\documentclass[11pt,epsf]{article}

\usepackage{epsfig}
\usepackage{amssymb}
\usepackage{graphicx}
\usepackage{color}
\usepackage{subfigure}

\begin{document}

\baselineskip 6mm
\renewcommand{\thefootnote}{\fnsymbol{footnote}}


\newcommand{\nc}{\newcommand}
\newcommand{\rnc}{\renewcommand}

\input amssym.def
\input amssym.tex


\newcommand{\tcb}{\textcolor{blue}}
\newcommand{\tcr}{\textcolor{red}}
\newcommand{\tcg}{\textcolor{green}}


\def\be{\begin{equation}}
\def\ee{\end{equation}}
\def\ba{\begin{array}}
\def\ea{\end{array}}
\def\bea{\begin{eqnarray}}
\def\eea{\end{eqnarray}}
\def\nn{\nonumber\\}


\def\ct{\cite}
\def\la{\label}
\def\eq#1{(\ref{#1})}


\def\a{\alpha}
\def\b{\beta}
\def\g{\gamma}
\def\G{\Gamma}
\def\d{\delta}
\def\D{\Delta}
\def\e{\epsilon}
\def\et{\eta}
\def\ph{\phi}
\def\Ph{\Phi}
\def\ps{\psi}
\def\Ps{\Psi}
\def\k{\kappa}
\def\l{\lambda}
\def\L{\Lambda}
\def\m{\mu}
\def\n{\nu}
\def\th{\theta}
\def\Th{\Theta}
\def\r{\rho}
\def\s{\sigma}
\def\S{\Sigma}
\def\ta{\tau}
\def\o{\omega}
\def\O{\Omega}
\def\z{\zeta}
\def\Z{\Zeta}

\def\pr{\prime}


\def\half{\frac{1}{2}}

\def\goto{\rightarrow}

\def\na{\nabla}
\def\grad{\nabla}
\def\curl{\nabla\times}
\def\div{\nabla\cdot}
\def\pa{\partial}

\def\bra{\left\langle}
\def\ket{\right\rangle}
\def\lb{\left[}
\def\lc{\left\{}
\def\ls{\left(}
\def\lp{\left.}
\def\rp{\right.}
\def\rb{\right]}
\def\rc{\right\}}
\def\rs{\right)}
\def\fr{\frac}
\def\inf{\infty}

\def\vac#1{\mid #1 \rangle}


\def\td#1{\tilde{#1}}
\def\check{ \maltese {\bf Check!}}
\def\text#1{{\rm #1}}

\def\Tr{{\rm Tr}\,}
\def\det{{\rm det}}


\def\bc#1{\nnindent {\bf $\bullet$ #1} \\ }
\def\ch {$<Check!>$ }
\def\ss {\vspace{1.5cm}}

\begin{titlepage}

\hfill\parbox{5cm} { }

\vspace{25mm}

\begin{center}
{\Large \bf Holographic renormalization in dense medium}

\vskip 1. cm
  {Chanyong Park\footnote{e-mail : cyong21@ewha.ac.kr}}

\vskip 0.5cm

{\it Institute for the Early Universe, Ewha Womans University, Daehyun 11-1, Seoul 120-750, Korea}
\\

\end{center}

\thispagestyle{empty}

\vskip2cm


\centerline{\bf ABSTRACT} \vskip 4mm

\vspace{1cm}

We investigate the holographic renormalization of a charged black brane 
with or without a dilaton field, whose dual field theory describes a dense medium at finite temperature. 
In a dense medium, two different thermodynamic descriptions are possible due to an additional conserved charge.
These two different thermodynamic ensembles are classified by 
the asymptotic boundary condition of the bulk gauge field. 
We show that in the holographic renormalization 
regularity of all bulk fields can reproduce consistent
thermodynamic quantities and that the Bekenstein-Hawking entropy is nothing but
the renormalized thermal entropy of the dual field theory.
Furthermore, we find that the Reissner-Nordstr\"{o}m AdS black brane is dual to a theory
with conformal matter as expected, whereas a charged black brane with 
a nontrivial dilaton profile is mapped to a theory with non-conformal matter 
although its leading asymptotic geometry still remains as AdS space.

\vspace{2cm}


\end{titlepage}

\renewcommand{\thefootnote}{\arabic{footnote}}
\setcounter{footnote}{0}


\section{Introduction}

The AdS/CFT correspondence \cite{Maldacena:1997re,Gubser:1998bc,Witten:1998qj,Witten:1998zw,Balasubramanian:1999jd} is a fascinating tool in studying a strongly interacting system like
quantum chromodynamics (QCD) or condensed matter system. Although there is no direct proof  
for the AdS/CFT
correspondence, it is conceptually supported from the same symmetry between a $d+2$-dimensional Anti de Sitter (AdS) gravity and 
$d+1$-dimensional conformal field theory (CFT). 
The key relation for this duality is 
that the semi-classical partition function of AdS gravity corresponds to the generating functional 
of a dual CFT \cite{Maldacena:1997re}
\be
{e^{- S_{grav}}} = \bra e^{- S} \ket_{CFT} .
\ee
This relation has been widely used and checked in various works related to integrability and localization \cite{Pestun:2009nn,Minahan:2002ve,Beisert:2003yb}.
For more realistic systems, it is important to encode the matter effect properly
which has been considered by considering the dual geometry in 
\cite{Lee:2009bya,Park:2009nb,Jo:2009xr}. 
In this paper, we will investigate such a matter theory by using the holographic renormalization technique \cite{Brown:1992br,Balasubramanian:1999re,Henningson:1998gx,Henningson:1998ey,Akhmedov:1998vf,de Haro:2000xn,Bianchi:2001de,Bianchi:2001kw,Skenderis:2002wp,Papadimitriou:2005ii,Akhmedov:2010mz}.

In general, the partition function of a quantum field theory suffers from the UV divergence which should
be renormalized by adding appropriate counter terms. This is also true in the dual gravity 
where the corresponding IR divergence is caused from infinite volume of the background geometry. In \cite{Brown:1992br,Balasubramanian:1999re}, 
it was shown that
adding local counter terms to the AdS boundary makes the on-shell gravity action finite and that
quantities derived from it can be reinterpreted as 
those of the dual CFT.
This holographic renormalization has been generalized even to the non-AdS geometries 
like the Lifshitz geometry \cite{Balasubramanian:2009rx,Korovin:2013nha,Ross:2009ar,Korovin:2013bua,Park:2013goa,Park:2013dqa} and Einstein-dilaton gravity \cite{Goldstein:2009cv,Charmousis:2010zz,Cadoni:2011nq,Goldstein:2010aw,Dong:2012se,Huijse:2011ef,Kanitscheider:2009as,Gouteraux:2011qh,Kulkarni:2012re,Park:2012cu,Kulkarni:2012in}. In these non-AdS examples the renormalized gravity action, after assuming the gauge/gravity duality,
was used to understand thermodynamics and hydrodynamics of the dual
non-relativisitic or non-conformal theory. In this procedure the counter terms can not only get rid of the UV divergence
but also yield additional finite contributions, so finding the correct counter terms is crucial for 
obtaining correct physical quantities.
This is also true in the renormalization of a quantum field theory.

In holography, a conformal field theory matches to a asymptotic AdS geometry. 
Introducing matter may break the conformal symmetry depending on the properties of matter. 
In the strong coupling regime, it is hard to investigate such matter effects
through the traditional quantum field theory method.  
Although there still remain many things to be clarified,
recent numerous works based on the AdS/CFT correspondence
provided some clues for understanding a strongly interacting system. 
In the holographic set-up, matter of the dual CFT can be realized as a field in AdS space.
Especially, a massless vector field in AdS is mapped to quark or hadronic number operator in the 
deconfining or confining phase respectively \cite{Lee:2009bya,Park:2009nb,Jo:2009xr,Park:2011zp,Lee:2013oya}.
Note that a black brane geometry is dual to a deconfining phase while
the non-black brane geometries, thermal AdS and thermal charged AdS 
in the hard wall model, describe a confining phase.    
From now on, we concentrate on a charged black brane geometry.

The holographic renormalization of the $d+2$-dimensional gravity theory leads to
the stress tensor of the $d+1$-dimensional boundary field theory.  
For a charged black brane, the dual theory contains quark matter.
In the thermodynamic point of view, the existence of matter makes two different thermodynamic interpretations possible 
\cite{Chamblin:1999hg,Emparan:1999pm}. 
On the gravity side, these two
different ensembles can be represented by the asymptotic boundary condition of the vector field. For example,
the grand canonical ensemble described by the chemical potential corresponds to the Dirichlet boundary
condition while the Neumann boundary condition is dual to the canonical ensemble. These two thermodynamic
systems are connected to each other through the Legendre transformation. 
In a charged black brane, there exists
a natural boundary condition for a vector field at the horizon. The regularity of a vector field
at the horizon is related to the Legendre transformation. As mentioned before, the holographic renormalization
requires appropriate counter therms. In addition, in the canonical ensemble
one more boundary term, the so called Neumannizing term, is needed to describe the
Neumann boundary condition which 
is crucial to obtain the correct thermodynamics. We show 
that physical quantities derived by the holographic renormalization perfectly 
satisfy the thermodynamic law in both thermodynamic ensembles.
Intriguingly, in the extremal limit of a Reissner-Nordstr\"{o}m AdS (RNAdS) black brane we find that 
the Fermi surface energy of the dual field theory is proportional to $N^{1/3}$ in the UV limit where $N$ denotes the number of quark. 
If there exists asymptotic freedom in the dual CFT,
one can identify the matter with the free relativistic fermion because it gives rise to $\e_F \sim N^{1/3}$.
However, since there is no asymptotic freedom in the holographic dual CFT, one 
can not directly interpret it as free relativistic fermions. We also consider another charged black brane
in the Einstein-Maxwell-dilaton theory in which the holographic renormalization also leads to the consistent 
physical quantities satisfying the thermodynamic law. Although the extremal limit is not well
defined in this case, one can still think of the Fermi surface in the low temperature limit.  
The Fermi surface energy of the Einstein-Maxwell-dilaton theory is also proportional to $\e_F \sim N^{1/3}$ 
similar to the RNAdS black brane case. The holographic renormalization results show that
dual matter of the RNAdS black brane does not break the conformal symmetry at the UV fixed point whereas
dual matter of the Einstein-Maxwell-dilaton theory breaks the conformal symmetry
in spite of the asymptotic AdS geometry.

The rest of paper is organized as follows. In Sec. 2, after applying the holographic renormalization 
to the Einstein-Maxwell theory, we calculate the on-shell actions and boundary stress tensors 
of two thermodynamic ensembles. From these result, we derive various thermodynamic quantities 
and show that they satisfy the thermodynamic law.
In Sec. 3, the holographic renormalization technique is applied to another asymptotic charged AdS black brane
of the Einstein-Maxwell-dilaton theory. Finally, we finish this work with some concluding remarks.


\section{Holographic renormalization of the RNAdS black brane}

Let us first take into account a 5-dimensional RNAdS black brane. 
A charged black brane solution usually has two hairs, charge and mass, so that its thermodynamics
can be represented as two different thermodynamic systems depending on the choice of the fundamental
thermodynamic variables \cite{Chamblin:1999hg}. On the gravity theory side, the choice of ensemble
is closely related to the asymptotic boundary condition of a bulk gauge field. 
For example, imposing the Dirichlet boundary condition on the gauge field corresponds to
taking a grand canonical ensemble, whereas one should impose the Neumann boundary condition in order to 
describe a canonical ensemble.  
In general, those two thermodynamic systems 
are related to each other by the Legendre transformation (see the details in Appendix A). 
According to the AdS/CFT correspondence, the black brane thermodynamics can be identified
with that of the dual field theory \cite{Balasubramanian:1999jd,Balasubramanian:1999re,Ross:2009ar,Korovin:2013bua,Park:2013goa,Charmousis:2010zz,Cadoni:2011nq,Park:2012cu,Emparan:1999pm}.
In this section, by using the holographic renormalization
we will investigate thermodynamics of the field theory dual to the RNAdS black brane. As will
be shown, the holographic renormalization results are perfectly matched to the black brane
thermodynamics.

For the direct comparison with the black brane thermodynamics, 
we regard an Euclidean Einstein-Maxwell action with a negative cosmological constant.
To obtain the Euclidean action from the Lorentzian one in \eq{act:lorentz},
one can apply  Wick rotation which describes rotation of time coordinate in a complex plane 
\be
x^t \to - i x^{\ta} \quad  {\rm and} \quad x_t \to  i x_{\ta} .
\ee
Under this Wick rotation, the field strength $F_{rt}$ with a Lorentzian signature 
transforms to
\be
F_{rt} \to \pa_r A_{t} - i \pa_{\ta} A_r . 
\ee
In order to preserve the usual definition form of the field strength $F_{MN} = \pa_{M} A_{N}
- \pa_{N} A_{M}$ up to an overall factor, 
we should also rotate the time component of the gauge field as $A_t \to i  A_{\ta}$.
Then, the Lorentzian field strength changes to the Euclidean one, 
$i F_{r \ta}$, with $A_M = \lc A_{\ta}, \vec{A} \rc$.
These Wick rotations lead to the following Euclidean action
\be			\la{act:einsteingmaxwellbulk}
S_E = \int d^5 x \sqrt{g} \lb - \fr{1}{2 \k^2}  \ls {\cal R} - 2 \L \rs + \fr{1}{4 g^2} F_{MN} F^{MN} \rb  ,
\ee
where the Euclidean metric is used. Since the Euclidean action
is the same as the Lorentzian one up to an overall minus sign, 
all equations of motion are same only except the fact that the Euclidean
metric should be used. For example, the Euclidean Einstein and Maxwell equations have the same forms
as the Lorentzian ones in \eq{eq:Einstein} and \eq{eq:Maxwell},
which allows the following Euclidean RNAdS black brane solution \cite{Lee:2009bya}
\bea
A_{\ta} &=& - i A_t = - i \ls g^2 \m - \fr{Q}{2 r^2} \rs , \nn
ds^2 &=& r^{2} f(r) \ d \ta^2 + \frac{dr^2}{r^2 f(r)} + r^2 (dx^2 + dy^2 + dz^2)  ,
\eea
with
\bea
f(r) =  1 - \fr{m}{r^4} + \fr{\k^2}{6 g^2} \fr{Q^2}{r^6} ,
\eea
where $m$ and $\m$ are a black brane mass and chemical potential respectively and 
we define its electric charge as $\sqrt{\fr{\k^2}{6 g^2}} Q$ for later convenience.
As mentioned before,
a charged black brane has two thermodynamic interpretations due to the conserved
charge. In the thermodynamic point of view, one corresponds to the grand canonical ensemble described by temperature, volume and chemical potential. 
If one use the charge as a fundamental variable
instead of the chemical potential, the black brane thermodynamics must be represented
as the canonical ensemble. 

The requirement of the regular metric at the horizon generally leads to a specific Euclidean time 
periodicity $\b$ corresponding to the inverse Hawking temperature. 
Due to the existence of the well-defined temperature, one can expect that the black brane
may be regarded as a thermal system. This is true if the entropy is given by the 
Bekenstein-Hawking entropy proportional to the horizon area. This thermodynamic
interpretation causes a big issue, why the entropy is proportional to the area but not
volume. There have been many attempts to account for this area law. One
of them is the holographic principle in which people have tried to map the black brane
thermodynamics to that of the boundary theory. Recently, the concept of the holography
has been further developed to the gauge/gravity duality of the string theory and
becomes a fascinating tool in studying a strongly interacting system 
\cite{Maldacena:1997re,Gubser:1998bc,Witten:1998qj,Witten:1998zw,Balasubramanian:1999jd,Skenderis:2002wp}. 
Following this holographic interpretation the area related to 
the Bekenstein-Hawking entropy is identified with the volume of the boundary theory where
the Bekenstein-Hawking entropy behaves as an extensive quantity expected in
ordinary thermodynamics. However, even in this case it is still mysterious why the Bekenstein-Hawking
entropy is defined at the horizon not the boundary. 
In general, the Hawking temperature is interpreted as temperature detected by an observer
living at the asymptotic boundary. Similarly other conserved quantities are also defined at the asymptotic boundary. 
In order to get well-defined thermodynamics of the boundary theory, it must be understood why the Bekenstein-Hawking
entropy appears as the thermal entropy of the dual theory.
The holographic renormalization together with natural assumptions,
as will be shown, indicates that the renormalized thermal entropy of the boundary theory is exactly given by
the Bekenstein-Hawking entropy.

Physically, it is quite natural to assume that all fields are regular in the entire space. If not so,
the theory becomes singular. 
In the Einstein-Maxwell theory, there exist two bulk fields, metric and vector fields.
At the horizon, the absence of a conical singularity in the metric determines the 
time periodicity. The Hawking temperature is represented as the inverse of
the time periodicity
\be		\la{const:HawkTemp}
T_H = \fr{r_h }{\pi} -  \fr{\k^2 Q^2}{12 \pi g^2} \fr{1}{r_h^5} ,
\ee 
where the horizon $r_h$ satisfies $f(r_h) = 0$.
Although the Hawking temperature is evaluated at the horizon, as mentioned before, it should be regarded
as temperature at the asymptotic boundary.
Under the coordinate transformation, the invariant combination of the vector field 
is given by $g^{\m\n}  A_{\m} A_{\n}$ which reduces to $g^{\ta\ta} A_{\ta}  A_{\ta}$ 
on the RNAdS black brane geometry. Since $g^{\ta\ta}$ diverges at the horizon, 
the regularity requires that $A_{\ta}$ should vanish at the horizon. This fact yields a relation between $\m$ and $Q$
\be     \la{const:regA}
\m  = \fr{Q}{2 g^2 r_h^2} .
\ee
In the holographic renormalization, these two consequences of natural assumptions 
can derive all desired thermodynamic quantities as will be shown.
It is worth to note that two parameters of the black brane, $m$ and $Q$, can
be represented as functions of $T_H$ and $\m$ (or $Q$ depending the boundary
condition of the vector field) \cite{Lee:2009bya}. From now on, we take into account  $T_H$ and $\m$ (or $Q$)
as fundamental parameters in which $r_h$ becomes a function of them.

In general, the gravity action without additional boundary terms
is not well defined because its variation becomes problematic at the boundary. Therefore,
the following two boundary terms should be added 
\be			\la{act:boundary}
S_{b} = \fr{1}{\k^2} \int_{\cal \pa M} d^d x  \sqrt{\g} \ \Th 
+ \fr{\z}{g^2} \int_{\pa {\cal M}} d^4 x  \sqrt{\g} \  n^r  F_{r \ta} \ A_{\ta} .
\ee
where $\g_{ab}$ and $\Th$ denote the induced metric and the extrinsic curvature. 
The unit normal vector is given by $n^r = \sqrt{g^{rr}}$. 
The first is the Gibbons-Hawking term and the last is called 
the Neumannizing term which determines the boundary condition of $A_{\ta}$,
e.g., $\z=0$ or $1$ for the Dirichlet or Neumann boundary condition respectively \cite{Balasubramanian:2009rx}.
As a consequence, the action having the well defined variations is given by the sum of the bulk in \eq{act:einsteingmaxwellbulk}
and boundary actions in \eq{act:boundary}.

The useful and crucial relation in the AdS/CFT correspondence
is that the on-shell gravity action is proportional to the free energy 
(or the generating functional) of the boundary (or dual) field theory
\be
 e^{- S_{on}} = \bra e^{- S} \ket_{CFT} = e^{- \b F},
\ee
where $F$ means a boundary free energy \cite{Maldacena:1997re}.
Despite of the previous regularity assumption, at the asymptotic boundary the on-shell gravity
action still suffers from 
some divergences caused by infinity in $r$. Following the holographic principle, these 
divergences
can be reinterpreted as the UV divergences of the dual field theory and removed by appropriate 
local counter terms analogous to quantum field theory. This method is called the holographic 
renormalization  \cite{Balasubramanian:1999re}. In this procedure counter terms not only remove the UV divergences but also provide finite contributions. Since physics crucially depends on these finite contributions
finding the correct counter terms is really an important issue. 

The correct counter terms should satisfy the following prescriptions
\begin{itemize}
\item The on-shell action should be finite.
\item The variation of the on-shell action with respect to the leading boundary value of all bulk fields
should also be finite.
\end{itemize}
As mentioned before, the first prescription corresponds to finiteness of the free energy. 
The second implies the physical quantities of the dual theory should be also finite.
In order to understand this, it should be noticed that in general the bulk fields 
denoted by $\Ph$ collectively have the following expansion form at the asymptotic boundary 
\cite{Gubser:1998bc,Witten:1998qj,Witten:1998zw}
\be
\Ph = J  \ r^{m_+} \ls 1 + \dots \rs +  \bra \bar{\cal O} \ket \ r^{m_-} \ls 1 + \dots \rs ,
\ee   
where ellipsis denotes small corrections. If $m_+>m_-$, on the dual field theory side,
$J $ and $\bar{\cal O}$ corresponds to a source and vev of its dual density operator respectively
and $\bar{\cal O}$ becomes a conjugate variable of $J $.
The on-shell gravity action, $S_{on}$, then includes the term related to the generating functional
\be
S_{on} = \cdots +  \int d \ta \int d^3 x \ J  \bra \bar{\cal O} \ket  + \cdots ,
\ee
where the Euclidean time integral runs from $0$ to $\b$.
Defining $\bra  {\cal O} \ket= \int d^3 x \bra  \bar{\cal O} \ket $,
the one point correlation function, when the source $J$ is uniform, is given by
\be
\b \bra  {\cal O}  \ket = - \fr{\pa S_{on} }{\pa J} ,
\ee
where the inverse temperature, $\b$, comes from the Euclidean time integral.
Therefore, the second prescription implies the one point function obtained by the holographic
renormalization should be finite
\be
\bra  {\cal O}  \ket = - \fr{1}{\b} \fr{\pa S_{on}}{\pa J} = {\rm finite} .
\ee
It is consistent with our intuition. Since $\bra  {\cal O}  \ket$ is a physical observable, it must be finite.
In the Einstein-Maxwell theory
there are two bulk fields, so the second prescription leads to
\be		\la{rel:onepointfn}
\bra T_{ab} \ket  = - \fr{2}{\b} \fr{\pa S_{on}}{\pa \g^{ab}} \quad {\rm and} \quad 
\bra N \ket  = - \fr{1}{\b} \fr{\pa S_{on}}{\pa \m } ,
\ee
where the factor $2$ is introduced for later convenience.
In the AdS/CFT correspondence, the electric charge of the RNAdS black brane is related to the 
quark number density operator of the dual field theory. So 
the quark number operator becomes $N = Q V_3$ where $V_3$ is the regularized volume at the asymptotic 
boundary.

\subsection{Grand canonical ensemble at the UV fixed point}

In the charged black brane geometry, the regularity conditions at the horizon in \eq{const:HawkTemp} and 
\eq{const:regA} determine two integration constants of the Einstein and Maxwell equations. 
After requiring the asymptotic AdS geometry, however,
there still remains one integration constant in the vector field 
which should be fixed by another boundary condition. One can usually choose a Dirichlet or Neumann
boundary condition at the asymptotic boundary.  

Let us first impose the Dirichlet boundary condition on the gauge field,
which fixes the chemical potential corresponding to the boundary value of the gauge field. 
Then, the dual theory is described by the grand potential of a grand canonical ensemble.
In this case, since the variation of the renormalized action with respect to the gauge field
is well defined, an additional boundary term is not needed. 
When the Neumann boundary condition instead of the
Dirichlet boundary condition is imposed,
this is not true anymore as will be shown in the next section.
The resulting renormalized action with the Dirichlet boundary condition becomes
\be
S_{ren}^{grand} = S_E + S_{GH} + S_{ct} ,
\ee
where $S_{ct}$ means the counter terms.
For the asymptotic AdS space, the correct counter terms have been
found in \cite{Balasubramanian:1999re}. In the Poincare patch of a five-dimensional AdS space, 
the counter term is given by
\be		\la{res:countterm}
S_{ct} = \fr{ 3}{\k^2} \int_{\pa {\cal M}} d^4 x \sqrt{\g} .
\ee
In the RNAdS black brane the vector field depending only on the radial coordinate
does not generate a new divergence, so an additional counter term is not required.

From the renormalized action, one can easily read the thermodynamic quantities by evaluating the boundary 
stress tensor.
At the UV fixed point ($r_0 \to \infty$), 
the grand potential from the on-shell gravity action yields
\bea		\la{res:holograndpot}
\O &\equiv& \lim_{r_0 \to \infty} \fr{S_{ren}^{grand} }{\b} , \nn
&=&  - \fr{V_3}{2 \k^2} r_h^4 - \fr{1}{3}  g^2  \m^2 \ V_3  r_h^2 ,
\eea
where \eq{const:regA} is used. 
This grand potential is consistent with the result obtained from the black brane thermodynamics 
\eq{res:grandcanres}.
The renormalized boundary energy-momentum tensor is then defined as \cite{Brown:1992br}
\bea
\bra {T^{\m}}_{\n} \ket &\equiv&  
\lim_{r_0 \to \infty}  \ls - \fr{2  \g^{\m\r}}{\b} \fr{\d S_{ren}^{grand}}{\d \g^{\r\n}} \rs \nn
&=& \lim_{r_0 \to \infty}  \ls \fr{1}{\k^2} \int d^3 x  \sqrt{\g} \ \ls {\Th^{\m}}_{\n} - {\d^{\m}}_{ \n} \Th \rs
 - 2  \g^{\m\r} \fr{\d S_{ct}}{\d \g^{\r\n}}  \rs ,
\eea
where the integral implies the boundary spatial volume integral.
The explicit internal energy and pressure read
\bea			\la{res:holoep}
E &=& {T^{0}}_{0} =   \fr{3 V_3}{2 \k^2} r_h^4  + g^2 V_3 \m^2   r_h^2 , \nn
P &=& - \fr{{T^{i}}_{i}}{V_3} = \fr{1}{2 \k^2} r_h^4 + \fr{1}{3} g^2   \m^2 r_h^2  ,
\eea
where $i$ is not summed.
From the previous second prescription the vev of the number operator becomes
\be		\la{res:numberoneptfn}
\bra N \ket \equiv   \lim_{r_0 \to \infty} \ls - \fr{1}{\b} \fr{\pa S_{ren}^{grand} }{\pa \m} \rs =
2 g^2 V_3 r_h^2 \ \m \, .
\ee
When imposing the Dirichlet boundary condition on the vector field, all physical quantities 
must be represented as functions of $T_H$, $V_3$, and $\m$ which are fundamental
variables describing the dual field theory. Since the grand potential of this system is
also a function of those variables, the following identity naturally appears
\be		\la{rel:exactdiffeq}
d \O  = \fr{\pa \O}{\pa T_H} d T_H + \fr{\pa \O}{\pa \m} d \m 
+ \fr{\pa \O}{\pa V_3} d V_3 .
\ee 
Since the grand potential in \eq{res:holograndpot} is linearly proportional to $V_3$,
its derivative with respect to $V_3$ satisfies 
\be
\fr{\pa \O}{\pa V_3} = - P .
\ee
Comparing $\pa \O / \pa \m$ with \eq{res:numberoneptfn}, we can easily see 
\be
\fr{\pa \O}{\pa \m} = - N ,
\ee
where $\bra N \ket$ is denoted by $N$. These relations indicate that the conjugate
variables of $V_3$ and $\m$ are given by $P$ and $N$ respectively. Now, let us define
\be
S \equiv - \fr{\pa \O}{\pa T_H} ,
\ee
then $S$ becomes the conjugate variable of $T_H$.
If the system we consider is a thermal system, $S$ can be naturally identified with the thermal 
entropy. Before checking the first law of thermodynamics, however, we can not say yet that our system is in thermal equilibrium. Using the fact that $r_h$ is given by a function of $T_H$ and
$\m$, $S$ at fixed $\m$ and $V_3$ reduces to 
\be
S  = \lp -  \fr{\pa \O}{\pa T_H} \right|_{\m,V_3} =
\fr{2 \pi V_3 }{\k^2} \ r_h^3  .
\ee
The values obtained here automatically satisfy the following relation
\be
\O  = E - T_H S  - \m N ,
\ee
which together with the exact differential in \eq{rel:exactdiffeq} lead to the first law of thermodynamics
\be
d E = T_H  dS + \m d N - P dV_3 .
\ee
These results show that the dual field theory is really a thermal system, as one expected.
Since the grand potential is the renormalized quantity of the dual field theory, 
the derived $S$ from it is also the renormalized one. 
As shown before, it satisfies all thermodynamic relations so that it 
should be identified with the thermal entropy of the dual field theory. Intriguingly, the thermal entropy derived here
is exactly the same as the Bekenstein-Hawking entropy in \eq{res:entropy}. As a result,
the holographic renormalization only with the quite natural regularity conditions leads to 
thermodynamics of the dual theory which is coincident with the black brane thermodynamics.
In this case, all thermodynamic quantities have appropriate dual theory interpretations. Especially,
the Bekenstein-Hawking entropy appears as the renormalized thermal entropy.

Note that the chemical potential $\m$  and
quark number operator $N=Q V_3$ are conjugate variables and, following the AdS/CFT correspondence, 
$\m$ is a source of $N$ (or $Q$) in the dual field theory.
Since the grand potential is described by the chemical potential, one can
easily evaluate the correlation function of the number operator in the grand canonical ensemble.
At the UV fixed point ($r_0 \to \infty$), the one-point function of the number operator is given by
\eq{res:numberoneptfn}, which on the gravity side corresponds to the electric charge
of the black brane. One can easily check that this one point function coincides with
the regularity of the vector field in \eq{const:regA}. Moreover, as will be seen in the next section, it
is also related to the Legendre transformation of the thermodynamic system.
From \eq{const:HawkTemp} and \eq{const:regA}, $r_h$ can be rewritten in terms of $\m$ and $T_H$ as
\be
r_h = \fr{\pi T_H + \sqrt{ \pi^2 T_H ^2 + \fr{4}{3} \k^2 g^2 \m^2}}{2} .
\ee
Using this, the expectation value of the number operator can be rewritten, in terms of the fundamental
thermodynamic variables of the grand canonical ensemble, as
\be
\bra N \ket = \fr{g^2  V^3}{2} \m \ls  \pi T_H + \sqrt{\pi^2 T_H^2 + \fr{4}{3} \k^2 g^2 \m^2 } \rs^2 .
\ee
Especially, at zero temperature ($r_h = \fr{  \k g  \m}{\sqrt{3}}$) this correlator
simply reduces to
\bea
\bra N \ket &=& \frac{2}{3} g^4 V_3 \kappa ^2 \mu ^3 .
\eea
Interestingly, this zero temperature expectation value of the number operator shows that the Fermi surface energy ($\e_F = \m$), if it exists \cite{Liu:2009dm},  is given by
\be		\la{res:fersuren}
\e_F = \ls \fr{3}{2 \k^2 g^4}\rs^{1/3}   \ls \fr{N}{V_3} \rs^{1/3} ,
\ee
where $N$ denotes the expectation value of the number operator.
In the general statistical mechanics, the Fermi surface energy of a non-relativistic free fermions, 
due to the nonrelativistic dispersion relation $\e = \fr{p^2}{2 m}$, yields
\be
\e_F  \sim \ls \fr{N}{V_3} \rs^{2/3} .
\ee 
On the other hand, for the relativistic free fermions with the dispersion relation, $\e =  p$, the 
Fermi surface 
energy in the natural unit is given by
\be
\e_F = (3 \pi^2)^{1/3} \ls \fr{N}{V_3} \rs^{1/3} ,
\ee
which looks very similar to the dual field theory result in \eq{res:fersuren}.
If the holographic dual theory of the asymptotic AdS geometry has an asymptotic freedom,
one can easily identify the holographic dual matter with the relativistic free fermion.
Unfortunately, because of the absence of an asymptotic freedom in the dual theory of the asymptotic 
AdS space,
one can not naively interpret the holographic dual matter as free fermion. So it would be interesting to construct 
the dual geometry of the field theory with an asymptotic freedom and deserves to study it further. 
Although the dual matter of the RNAdS geometry is not a relativistic free fermion, it
shows the relativistic free fermion-like Fermi surface energy which might be related to physical property 
of the non-Fermi liquid \cite{Liu:2009dm}.

Before concluding this section, it is worth to note that the trace of the boundary stress tensor vanishes. 
Since the second terms in \eq{res:holoep} are contributions from the holographic matter,
this result says that the holographic dual matter of the RNAdS black brane geometry
is also conformal. This is not always true in the asymptotic AdS geometry. As will be
shown later, a different asymptotic charged AdS black brane solution shows that
its dual matter can break the boundary conformal symmetry in spite of the asymptotic AdS geometry.

\subsection{Canonical ensemble}

As shown in \cite{Lee:2009bya}, 
imposing the Neumann boundary condition instead of the Dirichlet boundary
condition corresponds to the Legendre transformation in the dual theory, which
describes a canonical ensemble instead of a grand canonical ensemble. Although
there is no required additional counter term similar to the grand canonical ensemble case, 
a new boundary term should be added to satisfy the Neumann boundary condition.
Varying the gauge kinetic term with respect to the gauge field leads to a 
Neumannizing term \cite{Balasubramanian:2009rx}
\be			\la{res:Neumannizing}
\d S_{N} = \fr{1}{g^2} \int_{\pa {\cal M}} d^4 x \ \sqrt{g} \  g^{rr} g^{\ta \ta}  F_{r \ta} \ \d A_{\ta} .
\ee
This term, when the Dirichlet boundary condition is imposed, automatically vanishes 
due to $\d A_{\ta} = 0$, whereas it does not automatically vanish when the Neumann 
boundary condition is imposed.
In order to get a well defined gauge field variation with the Neumann boundary condition, one should
add an additional boundary term
\be
S_{bd} = \fr{1}{g^2} \int_{\pa {\cal M}} d^4 x  \ A_{\m} J^{\m} ,
\ee
where  $A_{\m} = \lc A_{\ta}, 0, 0, 0 \rc$ and $J^{\m} = \lc i Q, 0, 0, 0 \rc$.
Then, the variation of this new boundary term together with \eq{res:Neumannizing} gives rise to
\bea
\d S &=& \d S_{N} + \d S_{bd} , \nn
&=& \fr{1}{g^2} \int_{\pa {\cal M}} d^4 x \ \ls 
\sqrt{g} \  g^{rr} g^{\ta \ta}  F_{r \ta} + i Q  \rs \ \d A_{\ta} .
\eea
The vanishing of the action variation leads to the following Neumann boundary condition 
\be
\sqrt{g}  \ g^{rr} g^{\ta \ta}  F_{r \ta} = -  i Q ,
\ee
which is nothing but the Euclidean version of the Nuemann boundary condition in \eq{cond:Neumann}.
Therefore, the resulting renormalized action for the canonical ensemble
becomes
\be
S_{ren}^{can} = S_E + S_{GH} + S_{bd}  + S_{ct} .
\ee
From this renormalized action, the free energy of the dual canonical ensemble is given by 
\bea
F &\equiv& \lim_{r_0 \to \infty} \fr{S_{ren}^{can} }{\b} , \nn
&=& - \fr{ V_3}{2 \k^2}  r_h^4  + \fr{5 N^2}{12 g^2  V_3} \fr{1}{r_h^2} , 
\eea
where $r_h$ should be regarded as a function of $T_H$ and $N$. 
Comparing the free energy with the grand potential in \eq{res:holograndpot} says that
on the dual field theory side
the new boundary term reduces to $S_{bd}/\b= \m N$ and that the free energy and the grand potential
are related by the Legendre transformation as they should do.
After evaluating the boundary stress tensor similar to the previous case,
one can easily read the  internal energy and pressure 
in terms of $N$ 
\bea
E &=& {T^{0}}_{0} 
\equiv \lim_{r_0 \to \infty}  \ls - \fr{2  \g^{0\r}}{\b} \fr{\d S_{ren}^{can}}{\d \g^{\r 0}} \rs , \nn 
&= &  \fr{3  V_3}{2 \k^2}  r_h^4 +  \fr{N^2}{4 g^2  V_3} \fr{1}{r_h^2} , \nn
P_i  &=& - \fr{{T^{i}}_{i}}{ V_3} \equiv  \lim_{r_0 \to \infty}  \ls  \fr{2  \g^{i \r}}{\b V_3} \fr{\d S_{ren}^{can}}{\d \g^{\r i}} \rs , \nn
&= & \fr{1}{2 \k^2}  r_h^4 + \fr{N^2}{12 g^2  V_3^2} \fr{1}{r_h^2} .
\eea
These results exactly coincide with the RNAdS black brane thermodynamics of the canonical ensemble
in \eq{res:canonicalres1} and \eq{res:canonicalres2}.

\section{Holographic nonconformal matter in the asymptotic AdS geometry}

One can also take into account another deformed charged black brane which still has an asymptotic AdS geometry.
One of the known examples is the deformed RNAdS black brane with two different Liouville scalar potentials, 
where the scalar field was identified with a dilaton because it controls the
physical coupling \cite{Gubser:2009qt,Davison:2013uha,Gubser:2012yb,Alishahiha:2012ad}. 
In general, the dilaton field is related to the running coupling of the dual field theory, so
the existence of the nontrivial dilaton field implies that the dual field theory is deviated from
the UV conformal fixed point if the dual operator of the dilaton is relevant. In the deformed
charged black brane, the dilaton is coupled to the gauge field dual to the matter field so that
this nontrivial dilaton coupling can change the property of the dual matter field.
By using the the holographic renormalization studied in the previous section,  we will investigate
the dual field theory of the deformed charged black brane and the property of the dual matter.

With the Lorentzian signature, the action describing the deformed charged black brane is
given by
\be
S = \fr{1}{2 \k^2} \int d^5 x \sqrt{-g} \lb {\cal R} - \fr{1}{4} e^{4 \ph} F_{\m\n}  F^{\m\n}
- 12 \pa_{\m} \ph   \pa^{\m} \ph  + \fr{1}{R^2} \ls 8 e^{2 \ph} + 4 e^{- 4 \ph} \rs    \rb ,
\ee
where $R$ denotes the AdS radius. From now on, we set $R=1$ for convenience.
Then, the known solutions are
\bea
d s^2 &=&  r^2 e^{2 A(r)} \ls - f(r) dt^2 + d \vec{x}^2 \rs + \fr{e^{2 B(r)}}{r^2 f(r)} dr^2 , \nn
A_{\m} d x^{m} &=&  A_t dt ,
\eea
with
\bea
A(r) &=& \fr{1}{3} \log \ls 1+ \fr{Q^2}{8 m r^2} \rs , \nn
B(r) &=& - \fr{2}{3} \log \ls 1+ \fr{Q^2}{8 m r^2} \rs , \nn
f(r) &=& 1 - \fr{m}{\ls r^2 + \fr{Q^2}{8 m } \rs^2} , \nn
A_t &=& 2 \k^2 \m -  \fr{Q}{ 2 \ls r^2 + \fr{Q^2}{8 m }\rs } , \nn
\ph(r) &=& \fr{1}{6} \log \ls 1+ \fr{Q^2}{8 m r^2} \rs ,
\eea
where we used slightly different notations from \cite{Gubser:2009qt} for more concrete thermodynamic
interpretation.
In the above, free parameters, $m$, $Q$ and $\m$, correspond to the mass,
charge and chemical potential of the charged black brane respectively. More precisely,
 if one expands the gauge 
field near the asymptotic boundary to
\be
A_t = 2 \k^2 \m - \fr{Q}{2 r^2} + {\cal O} \ls \fr{1}{r^4}\rs ,
\ee
two integration constants $\m$ and $Q$ appear. Furthermore, following the AdS/CFT correspondence these
values can be reinterpreted, similar to the previous case, as the quark chemical potential $\m$ and 
number density $Q$ 
\be
Q = \lim_{r \to 0}  \sqrt{-g} \ e^{4 \ph} g^{tt} g^{rr} \pa_r A_t     .
\ee 
Note that for $Q \to 0$, the above solution simply reduces to the usual Schwarzschild AdS black brane.

The absence of a conical singularity at the horizon gives rise to Hawking temperature 
which is associated with the horizon $r_h$
\be
r_h = \pi T_H   ,
\ee
and the Bekenstein-Hawking entropy leads to
\be			\la{res:dialBHentropy}
S_{BH} = \fr{2 \pi V_3}{\k^2} \ls r_h^3 + \fr{Q^2 }{8 m} r_h \rs .
\ee
From the vanishing of the black brane factor and the gauge potential at the horizon, 
one can easily find two relations between parameters
\bea
Q &=& 4 \k^2 \m \sqrt{m}  , \nn
\sqrt{m} &=& r_h^2 + 2 \k^4 \m^2  ,
\eea
which are convenient in describing the grand canonical ensemble because all other parameters are given by
functions of $T_H$ and $\m$. In the canonical ensemble, $m$ and $\m$ should be considered as
functions of $T_H$ and $Q$, which have complicated forms
\bea
\m &=& \fr{Q}{4 \k^2 \sqrt{m}} , \nn
\sqrt{m} &=& \frac{1}{3}  r_h^2 +
 \frac{ 2^{4/3} \ r_h^4}{3 \ls 16 r_h^6 + 27 Q^2 + 3 Q \sqrt{96 r_h^6+81 Q^2} \rs^{1/3}} \nn
&&  \quad + 
 \frac{\ls 16 r_h^6 + 27 Q^2 + 3 Q \sqrt{96 r_h^6+81 Q^2} \rs^{1/3}}{3 \ 2^{4/3}}  .
\eea
From the thermodynamic point of view, since $m$ is given by a very complicated function of $T_H$ and $Q$
in the canonical ensemble,
it is not easy to find  analytic thermodynamic quantities directly from the thermodynamic laws.
From now on, we try to evaluate the thermodynamic quantities through 
the holographic renormalization scheme instead of  the thermodynamic laws.

With the Euclidean signature, the action for the grand canonical ensemble
is
\be
S_{grand} = - \fr{1}{2 \k^2} \int d^5 x \sqrt{-g} \lb {\cal R} - \fr{1}{4} e^{4 \ph} F_{\m\n}  F^{\m\n}
- 12 \pa_{\m} \ph   \pa^{\m} \ph  + \fr{1}{R^2} \ls 8 e^{2 \ph} + 4 e^{- 4 \ph} \rs    \rb .
\ee
In the holographic renormalization, one should add the Gibbons-Hawking and counter terms in \eq{act:boundary}.  If we impose the Dirichlet boundary condition on $A_{\ta}$, the dual
thermodynamics represents the grand canonical ensemble in which the chemical potential 
instead of the number operator becomes a basic thermodynamic variable. Then, the grand potential 
proportional to the on-shell gravity action leads at the asymptotic boundary to
\bea
\O (T_H, \m, V_3) &\equiv& \fr{S_E}{\b} \nn
&=& -\frac{\pi ^4 V_3 }{2 \kappa ^2} T_H^4 -2 \pi ^2 \kappa ^2  V_3 T_H^2 \mu ^2 -\frac{10}{3}
   \kappa ^6  V_3  \mu ^4 .
\eea
Furthermore, from the boundary stress tensor the internal energy and pressure are represented
as functions of $T_H$ and $\m$
\bea
E &\equiv& T^0_0  \nn
&=& \frac{3 \pi ^4 V_3 }{2 \kappa ^2} T_H^4 + 6 \pi ^2 \kappa ^2  V_3 T_H^2 \mu ^2  +\frac{14}{3}
   \kappa ^6  V_3  \mu ^4 , \nn
P &\equiv& - \fr{T^i_i}{V_3}  = \nn
&=& \frac{\pi ^4  }{2 \kappa ^2} T_H^4 + 2 \pi ^2 \kappa ^2  T_H^2 \mu ^2 
+\frac{10}{3}
   \kappa ^6   \mu ^4 .
\eea
The one-point function of the number operator is
\be
N \equiv  \lim_{r_0 \to \infty} \ls - \fr{1}{\b} \fr{\pa S_{ren}^{grand} }{\pa \m} \rs =
\frac{4}{3} \kappa ^2 \mu  V_3 \left(3 \pi ^2 T_H^2+10 \kappa ^4 \mu ^2\right) .
\ee
Similar to the holographic renormalization of the RNAdS black brane, 
the renormalized thermal entropy $S$ appears as the conjugate variable of $T_H$ 
\be
S \equiv   \lp -  \fr{\pa \O}{\pa T_H} \right|_{\m,V_3} =
\frac{2 \pi ^2 V_3 T_H }{\kappa ^2} \left(\pi ^2 T_H^2+2 \kappa ^4 \mu ^2\right) 
\ee
and satisfies the thermodynamic relation
\be
\O = E - T_H S - \m N .
\ee
Comparing it to the Bekenstein-Hawking entropy in \eq{res:dialBHentropy}, 
one can easily see that the derived thermal entropy is exactly the same as the 
Bekenstein-Hawking entropy and all thermodynamic quantities derived here are
consistent with those obtained from the black brane geometry.

For $\m=0$, the thermodynamics of the Schwarzschild AdS black brane is reproduced.
When $Q = 2^{3/2} \ m^{3/4}$, the  charged black brane reaches to the extremal limit where
$T_H$ and $S_{BH}$ reduces to zero. It should be however noted that
the extremal limit has a naked singularity at the center ($r=0$) so that the zero temperature is not
well defined. In the very low temperature limit ($T_H \ll \k^2 \m$), the expectation value of the number
operator is proportional to $\m^3$. If there exists a Fermi surface,  
the Fermi surface energy leads to
\be
\e_F \sim N^{1/3} ,
\ee
which is similar to the RNAdS black brane case. 
Unlike the RNAdS black brane case, however, the trace of the boundary stress tensor shows a 
nonzero value depending on the chemical potential
\be
\bra {T^{a}}_{a} \ket_{\m} = E - 3 P V= - \fr{16}{3} \k^6 V_3 \m^4  ,
\ee
where the subscript $\m$ implies a nontrivial vacuum state $\left| 0 \ket_{\m}$.
This result indicates that the dual matter breaks the conformal symmetry. 
Why does not the trace of the boundary stress tensor vanish in spite of the asymptotic AdS geometry?
The reason is as follows. Although the deformation of the asymptotic AdS geometry by the dilaton
and gauge field does not change the leading behavior of the metric, it can contribute some 
subleading corrections to the metric at $ r^{- 2}$ order. 
Since these subleading terms provide nonzero finite contribution to both the on-shell action and 
boundary stress tensor even at the UV cutoff, the dual conformal symmetry is broken 
without changing of the leading asymptotic geometry.

In order to describe the canonical ensemble instead of the grand canonical ensemble,
we should impose the Neumann boundary condition on the vector field which requires one more boundary term 
\be		\la{act:GRcan}
S_{can} = S_{grand} + \fr{1}{g^2} \int_{\pa {\cal M}} d^4 x  \ A_{\m} J^{\m} ,
\ee
where $A_{\m} = \lc - i A_{t}, 0, 0, 0 \rc$ and $J^{\m} = \lc i Q, 0, 0, 0 \rc$.
Since the new additional boundary term cancels Neumannizing term, the vector field
variation is well defined. In the dual theory point of view, 
inserting this boundary term corresponds to the Legendre transformation
because the new boundary term reduces to
\be
 \fr{1}{g^2} \int_{\pa {\cal M}} d^4 x  \ A_{\m} J^{\m} = \b \m N .
\ee 
Therefore, \eq{act:GRcan} becomes
\be
F = \O + \m N ,
\ee
where $F \equiv S_{can}/\b$ and $\O \equiv S_{grand}/\b$ denote the free energy and grand potential
respectively. As mentioned before, it represents the Legendre transformation between the grand canonical
and canonical ensemble. 
The on-shell action and  boundary energy-momentum tensor in the canonical ensemble 
yield the following free energy, internal energy and pressure 
\bea        \la{res:thresEMdblackbr}
F &=& -\frac{\pi ^4 V_3 }{2 \kappa ^2} T_H^4 + 2 \pi ^2 \kappa ^2  V_3 T_H^2 \mu ^2+\frac{14}{3}
   \kappa ^6  V_3 \mu ^4 , \nn
E  &=& \frac{3 \pi ^4 V_3 }{2 \kappa ^2} T_H^4+ 6 \pi ^2 \kappa ^2  V_3 T_H^2 \mu ^2  +\frac{14}{3}
   \kappa ^6  V_3  \mu ^4 , \nn
P &=& \frac{\pi ^4  }{2 \kappa ^2} T_H^4 + 2 \pi ^2 \kappa ^2  T_H^2 \mu ^2 
+\frac{10}{3}
   \kappa ^6   \mu ^4 .
\eea
where, since the basic variables of the canonical ensemble are $T_H$, $N$ and $V_3$,
the chemical potential is given by a function of those variables
\be
\m = \frac{4 \ 3^{1/3} \pi ^2 \kappa ^6 T_H^2- 2^{1/3} \left(\sqrt{96 \pi ^6 \kappa
   ^{18} T_H^6+81 \kappa ^{16} \fr{N^2}{V_3^2}}-9 \kappa ^8 \fr{N}{V_3} \right){}^{2/3}}{2\ 6^{2/3} \kappa ^4
    \ls \sqrt{96 \pi ^6 \kappa ^{18} T_H^6+81 \kappa ^{16} \fr{N^2}{V_3^2}}-9 \kappa ^8 \fr{N}{V_3}\rs^{1/3} } .
\ee
The holographic renormalization results in \eq{res:thresEMdblackbr} coincide with the 
black brane thermodynamics and satisfies all thermodynamic relations.


\section{Discussion}

We have studied the holographic renormalization of the charged black branes whose
asymptotic geometry is given by an AdS space. In the general renormalization scheme of the field theory,
finding correct counter terms are very important because they cancel the UV divergences and at same
time provide some finite contributions. Therefore, the renormalized physical quantities usually depend 
on the finite contributions of the counter terms. This is also true in the holographic renormalization.  
Following the AdS/CFT correspondence and using the holographic renormalization, 
in this paper, we have investigate the physical quantities of the dual field theory with matter.
As mentioned before, for obtaining the correct physical results one should take into account the
correct counter terms. In the charged black brane with an asymptotic AdS geometry, 
the bulk gauge field dual to matter does not require a new divergent terms in the on-shell gravity action
so that one need to introduce more counter terms except one used in the Schwarzschild (neutral) AdS
black brane. Although there is no additional counter terms, the change of the metric caused by the
bulk gauge field leads to a new finite contribution to the on-shell gravity action which depends on
the physical quantities of matter, the chemical potential or number density. 

In the RNAdS black brane case, the boundary stress tensor generated by the on-shell gravity action
becomes traceless. As expected usually, the dual matter does not break the conformal symmetry
of the asymptotic AdS geometry. Furthermore, the expectation value of the number operator shows that
the Fermi surface energy at zero temperature is proportional to  $N^{1/3}$ which intriguingly is resemble to 
that of the free relativistic fermion. In spite of the resemblance, however, one can not identify the
dual matter of the RNAdS black brane with a free relativistic fermion due to the absence of the 
asymptotic freedom. So it is more preferable to identify the dual matter with the non-Fermi liquid, strongly
interacting fermion.

In the RNAdS black brane, there is no nontrivial dilaton field associated with the running coupling constant
of the dual field theory. We further regarded the charged black brane in the Einstein-Maxwell-dilaton theory.
The asymptotic geometry of it is also given by an AdS space, which is the leading part of the metric, and
the next to leading part gives rise to the nontrivial finite contribution to both the on-shell gravity action
and the boundary stress tensor. Unlike the RNAdS black brane, the boundary stress tensor of the 
charged dilatonic black brane is not traceless which implies that the dual matter breaks the conformal
symmetry although the leading part of the asymptotic metric is not changed.

Finally, we also showed that the holographic renormalization results can be reinterpreted
as two different ways depending on the asymptotic boundary condition of the gauge field. These result
are exactly matched with the thermodynamic laws of the charged black brane geometries we considered. 
All above results are derived at the UV fixed point. So it is interesting to investigate the renormalizaiton
group flow of them.

\vspace{1cm}

{\bf Acknowledgement}

We would like to thank Yun Soo Myung and Mu-in Park for valuable discussion.
This work has been supported by the WCU grant no. R32-10130 and the Research fund no. 1-2008-2935-001-2
by Ewha Womans University.  
C. Park was also supported by Basic Science Research Program through the National Research Foundation of 
Korea(NRF) funded by the Ministry of Education (Grant No. NRF-2013R1A1A2A10057490).

\vspace{1cm}

\appendix

\section{Thermodynamics of the RNAdS black brane}

\renewcommand{\theequation}{A.\arabic{equation}}
\setcounter{equation}{0}

Here, we will explain the thermodynamics of the RNAdS black brane in details
for the comparison with the holographic renormalization results. The action for the RNAdS black brane
with a Lorentzian signature is given by
\be \la{act:lorentz}
S_L = \int d^5 x \sqrt{-g} \lb \fr{1}{2 \k^2}  \ls {\cal R} - 2 \L \rs - \fr{1}{4 g^2} F_{MN} F^{MN}\rb   ,
\ee
where $\L = - 12/R^2$ and $R$ is the AdS radius.
From now on, we set $R$ to be $1$ for convenience.
The Einstein and Maxwell equations lead to
\bea
{\cal R}_{MN} -\half g_{MN} {\cal R} + g_{MN} \L &=&
\fr{\k^2}{g^2} \ls F_{MP} {F_N}^P - \fr{1}{4} g_{MN} F_{PQ} F^{PQ} \rs  , \la{eq:Einstein} \\
0 &=& \pa_M \ls \sqrt{-g} \ g^{MP} g^{NQ} F_{PQ} \rs \la{eq:Maxwell}  .
\eea
If we turn only on the time component of the gauge field $A_t$ as a function of the radial
coordinate $r$, the RNAdS black brane solution satisfying these two equations is given by
\bea
A_t &=& g^2 \m - \fr{Q}{2 r^2} , \la{sol:maxwell} \\
d s^2 &=& - r^2 f(r) \ dt^2 + \fr{1}{ r^2 f(r)} dr^2 + r^2\ \d_{ij} \ dx^i  d x^j  , \la{sol:Einstein} 
\eea
with
\bea
f(r) =  1 - \fr{m}{r^4} + \fr{\k^2}{6 g^2} \fr{Q^2}{r^6} ,
\eea
where $\m$ and $Q$ are two arbitrary integration constants describing the electric chemical
potential and charge density respectively. 
Imposing an appropriate asymptotic boundary condition on $A_t$ can fix one of two 
integration constants. There are two possibilities which provide explicit thermodynamic meanings to
constants.
One is the Dirichlet boundary condition which determines the 
boundary value of $A_t$, $\m$, and the other is the Neumann boundary condition in order to fix the charge $Q$. 

Let us first take into account the case with the Neumann boundary condition.
At the asymptotic boundary the Neumann boundary condition for the gauge field is given by 
$Q$
\be			\la{cond:Neumann}
Q = \lim_{r \to \inf} \sqrt{-g} \ g^{rr} g^{tt} F_{rt} ,
\ee
which fix the electric charge density as $Q$. Note that 
the chemical potential still remains as an undetermined integration constant.
Due to the vanishing of $f(r)$ at the event (or outer) horizon $r_h$, the black brane mass density can be
rewritten, in terms of $r_h$ and $Q$, as
\be
m = r_h^4  + \fr{\k^2}{6 g^2} \fr{Q^2}{r_h^2}
\ee
Using this, the Hawking temperature defined by the surface gravity at the horizon
reduces to
\be			\la{res:temperature}
T_H = \fr{1}{\pi} \ r_h -  \fr{\k^2 Q^2}{12 \pi g^2} \fr{1}{r_h^5} .
\ee
It is well known that a black brane (or hole) provides a well-defined thermodynamic system, so
one can regards, in the thermodynamic description, $r_h$ as a function of the other thermodynamic variables, 
$T_H$ and $Q$, from \eq{res:temperature}. The Bekenstein-Hawking entropy reads
\be			\la{res:entropy}
S_{BH} = \fr{2 \pi V_3 }{\k^2} \ r_h^3 ,
\ee
where $V_3$ is a regularized  $3$-dimensional  spatial volume. Introducing the total charge 
$N = Q V_3$, the RNAdS black brane geometry can be classified by $T_H$, $N$
and $V_3$, which are thermodynamic variables describing the free energy of the canonical
ensemble. From the thermodynamic relations together with \eq{res:temperature} and 
\eq{res:entropy}, one can easily derive other thermodynamic quantities like
the free energy $F$, internal energy $E$, and pressure $P$. 

In the canonical ensemble, the most important thermodynamic
function  is the free energy 
\be		\la{rel:freeencanon}
F  = E - T_H S_{BH}  ,
\ee
from which all thermodynamic quantities can be derived.
Applying the first law of thermodynamics, 
\be
dE = T_H d S_{BH} + \m d N - P d V_3 , 
\ee
the free energy of the canonical ensemble satisfies the following thermodynamic relation
\be			\la{res:secondthermo}
d F = - S_{BH} \ d T_H + \m d N - P d V_3 .
\ee
Since a RNAdS black brane solution in \eq{sol:Einstein} is described by $Q = N/V_3$,  
it can be easily reinterpreted as thermodynamics of a canonical ensemble. 
After rewriting the Hawking temperature as a function of the Bekenstein-Hawking entropy
with a fixed volume and total charge, the integral of the first thermodynamic law 
gives rise to
\be		\la{res:canonicalres1}
E = \int d S_{BH} \ T_H= \fr{3  V_3}{2 \k^2}  r_h^4 +  \fr{N^2}{4 g^2  V_3} \fr{1}{r_h^2} .
\ee 
In the above integration, it is also possible to add an arbitrary function depending only on $N$ and $V_3$
because including such a function still satisfies 
\be
T_H  =  \lp \fr{\pa E}{\pa S_{BH}} \right|_{N,V_3} .
\ee
However, this additional function does not satisfy the other thermodynamic relations so that 
such a term should vanish.
Then, \eq{rel:freeencanon} gives rise to the free energy
\be
F =   - \fr{ V_3}{2 \k^2}  r_h^4  + \fr{5 N^2}{12 g^2  V_3} \fr{1}{r_h^2} ,
\ee
and, when temperature and the total charge are fixed, its derivative with respect to the volume leads
to the pressure 
\be		\la{res:canonicalres2}
P = \lp \fr{\pa F}{\pa V_3} \right|_{T_H,N} =    \fr{1}{2 \k^2}  r_h^4 + \fr{N^2}{12 g^2  V_3^2} \fr{1}{r_h^2} .
\ee		
From the thermodynamic relation, the chemical potential of this system is given by a function of $N$
\be			\la{res:Legendretransf}
\m = \lp \fr{\pa F}{\pa N} \right|_{T_H,V_3}  = \fr{N}{2 g^2 V_3} \fr{1}{r_h^2} ,
\ee
which is nothing but the regularity condition for the time component of the gauge field
in \eq{const:regA}.
Note that $r_h$ in the canonical ensemble should be regarded as a function of $T_H$, $N$ and $V_3$.
The above results show that the equation of state parameter of the dual system is given by $1/3$ 
like a pure $AdS_5$ and Schwarzschild-type AdS black brane. Therefore, the dual matter of the RNAdS
black brane is conformal because the trace of the energy momentum tensor is zero, $E-3 P V_3 = 0$.

Now, let us consider the Dirichlet boundary condition instead of the Neumann boundary condition, which
is given at the asymptotic boundary by
\be
\m = \lim_{r \to \inf} \fr{A_t}{g^2} .
\ee
In the thermodynamic interpretation, the chemical potential is a fundamental variable of the grand potential
in the grand canonical ensemble. So imposing the Dirichlet boundary condition is associated to the description
of the grand canonical ensemble.  In this case, the grand potential
is given by a function of temperature, chemical potential and spatial volume, $\O (T_H,\m,V_3)$.
Moreover, the charge $N$ should be a function of $\m$ and $T_H$, which can be derived from 
$N= - \lp \fr{\pa \O}{\pa \m} \right|_{T_H, V_3}$. 
In the above RNAdS black brane geometry, all solutions are not represented by $\m$ but $Q$.
Therefore, one can not directly read the thermodynamics of the grand canonical ensemble without
any additional relation. 
Using the regularity of the $A_t$ norm at the horizon, $Q$ can be fixed as
a function of $\m$ 
\be		\la{res:chpotden}
Q = 2 \ g^2 \ r_h^2 \ \m ,
\ee
which is related to the Legendre transformation. In the grand 
canonical ensemble description, $r_h$ must be regarded as a function of $T_H$, $\m$ and $V_3$.
Similar to the previous case, the first thermodynamic law and the following relations
 \bea
\O  &=& E - T_H S_{BH}  - \m N , \\
d \O &=& - S_{BH} d T_H - N d \m - P d V_3 ,
\eea 
determines the grand potential as the following form 
\be		\la{res:grandcanres}
\O (T_H,\m,V_3) =  - \fr{V_3}{2 \k^2} r_h^4 - \fr{g^2  \m^2 \ V_3 }{3}  r_h^2 .
\ee 
As mentioned before, the grand potential is related to the free energy by the Legendre transformation
\bea
\O  &=& F - \m N .
\eea
The other interesting thermodynamic quantities read
\bea
E &=&  \fr{3 V_3}{2 \k^2} r_h^4  + g^2 V_3 \m^2   r_h^2 , \nn
P &=&  \fr{1}{2 \k^2} r_h^4 + \fr{1}{3} g^2   \m^2 r_h^2  .
\eea
which are equivalent to the results obtained in the canonical ensemble if
rewriting the chemical potential as a function of the charge by using \eq{res:Legendretransf}. \\ \\


\end{document}